# Effect of Lead Substitution on $LaO_{0.5}F_{0.5}BiS_2$

**S. Otsuki, S. Demura, Y. Sakai Y. Fujisawa and H. Sakata**

*Tokyo University of Science, Department of Physics, Shinjuku-ku, Tokyo 162-8601, Japan*

demura@rs.tus.ac.jp.

**Abstract**.
We examined Lead (Pb) Substitution effect on a single crystal of a layered superconductor $LaO_{0.5}F_{0.5}BiS_2$. Pb concentration dependence of the lattice constant showed slight anomaly at about 8% and 9% substitution of Pb for Bi. These samples showed the enhancement of the superconducting transition temperature and the superconducting volume fraction. Furthermore, these samples showed the anomaly in the temperature dependence of the resistivity at about 150K. These results were not observed in Pb substituted $NdO_{0.7}F_{0.3}BiS_2$. Therefore, the enhancement of the superconducting properties by Pb substitution is related to the structural instability for the pale perturbation in $LaO_{0.5}F_{0.5}BiS_2$.

## 1. Introduction

Recently, a number of new layered superconductors $Ln$(O,F)$BiS_2$ ($Ln$=La, Ce, Pr, Nd, Sm, Yb and Bi) have been reported [1-5]. These materials have a layered structure composed of alternative stacks of a block layer and a superconducting $BiS_2$ layer. Superconductivity in $Ln$O$BiS_2$ emerges when carriers are doped by substituting oxygen for fluorine [2]. Although the superconducting transition temperature ($T_c$) of these materials is about 4 K, $T_c$ has been enhanced by several techniques.

One is the application of hydrostatic pressure. For instance, La(O,F)$BiS_2$ shows a $T_c$ of approximately 3K at ambient pressure. The $T_c$ increases up to 10 K under high pressure, which is the highest $T_c$ in $BiS_2$-based superconductors [6,7]. Simultaneously, a superconducting shielding fraction also increases from 10% in ambient pressure to 100 % in high pressure. The enhancement of these superconducting properties is expected to attribute an existence of a high-pressure phase detected by X-ray diffraction measurement under high pressure, where the crystal structure changes from tetragonal to monoclinic [8]. Although the high-pressure phase returns to the low-$T_c$ phase with releasing the pressure, a high-pressure synthesis can keep the high-$T_c$ phase under ambient pressure: the sample synthesized under the high pressure shows higher $T_c$ under ambient pressure [9]. In addition, when an annealing under high pressure is performed to the sample prepared in ambient pressure, the annealed sample keeps the higher-$T_c$ phase under ambient pressure [2]. It is not known whether this high $T_c$ phase prepared by high pressure annealing or high pressure synthesis is corresponding to the phase appeared under high pressure measurement because the crystal structure of former samples has not been determined due to broadening of XRD peaks attributed to declining the crystallinity. However, these studies indicate that the existence of another stable atomic configuration with higher $T_c$ in $BiS_2$-based superconductors.

Another is a partial substitution of an element with different ionic radius. The substitution causes an strain in the crystals, and increases $T_c$, which is called as chemical-pressure effect [12-15]. For example, a partial substitution of Selenium (Se) with Sulfur (S) into the superconducting layer in La(O,F)$BiS_2$ enhances $T_c$ and superconducting volume fraction [10-11]. The partial substitution of lanthanide with different lanthanide also enhances $T_c$ [12-15]. These results were analyzed with an in-plane chemical pressure defined by Mizuguchi [16]. These results indicate that superconducting properties in $BiS_2$-based superconductors are sensitive to the subtle change in the crystal structure.

Recently, we performed the substitution of Pb for Bi in $NdO_{0.7}F_{0.3}BiS_2$. Since Pb ion has larger ionic radius than that of Bi ion, the substitution induces the chemical pressure into the sample. $T_c$ was successfully enhanced by the substitution of Pb for Bi [17]. Thus, it is interesting how the substitution affects the superconducting properties in $LaO_{0.5}F_{0.5}BiS_2$, which is more sensitive to the external perturbation such as external pressure than $NdO_{0.7}F_{0.3}BiS_2$.

Here, we examine the Pb-substitution effect on the superconducting properties in single crystal $LaO_{0.5}F_{0.5}BiS_2$. Pb concentration dependence of the lattice constant showed slight anomaly at about 8% and 9% substitution of Pb for Bi. These samples showed the enhancement of the superconducting transition temperature and the superconducting volume fraction. Furthermore, these samples showed the anomaly in the temperature dependence of the resistivity at about 150K. These anomalies observed in the lattice constant and electrical resistivity seem to stem from the structural instability against external perturbations in $LaO_{0.5}F_{0.5}BiS_2$. This fact opens the new insight about the enhancement of superconducting properties in $BiS_2$-based superconductors.

## 2. Experiment

Single crystalline samples of $LaO_{0.5}F_{0.5}Bi_{1-x}Pb_xS_2$($x$=0-0.10) were synthesized with a CsCl/KCl flux method in an evacuated quartz tubes 18]. Powders of $La_2S_3$, $Bi_2O_3$, $Bi_2S_3$, $PbF_2$ and $BiF_3$ with Bi grains are used as a starting material. The $Bi_2S_3$ powders were prepared by reacting Bi and S grains in an evacuated quartz tube at 500 ˚C for 10 hours. The mixture of staring materials and CsCl/KCl powder of 7.5g was sealed in the evacuated quartz tube. The tube was heated at 900˚C for 12 hours and kept it at 900 ˚C for 24 hours and cooled down to 630 ˚C at the rate of 0.5 ˚C/h or 1 ˚C/h. After this thermal process, the obtained material was washed by distilled water to remove the flux. Powder X-ray diffraction (PXRD) patterns were collected by a Rigaku X-ray diffractometer with Cu Kα radiation using $\theta$-2$\theta$ method. These powder samples used in PXRD measurements were prepared by grinding single crystals. Temperature dependence of the magnetic susceptibility down to 2 K was measured with MPMS (Magnetic Property Measurement System). Temperature dependence of the electrical resistivity was measured down to 2.5K with four terminals method.

## 3. **Result**

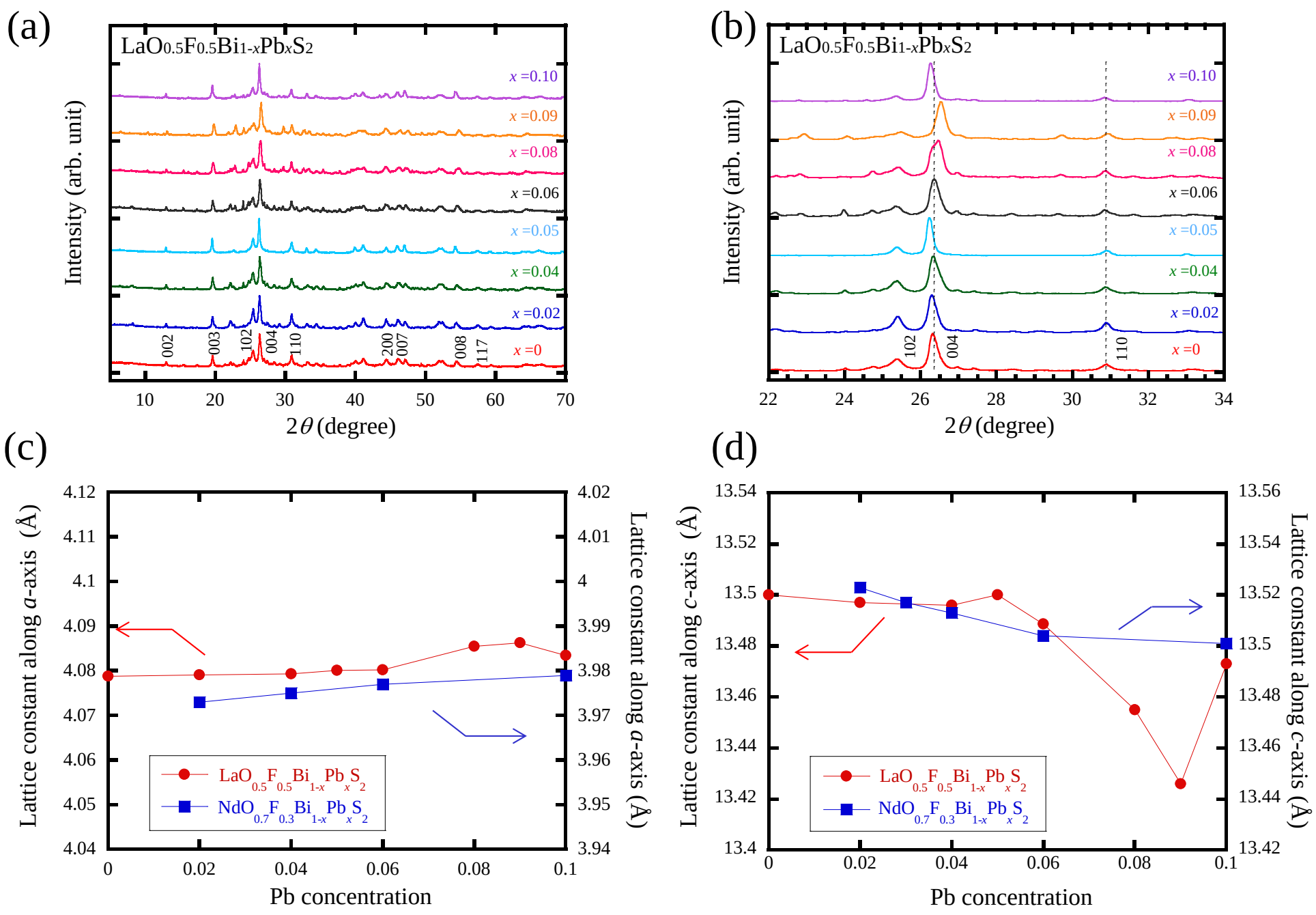

Figure 1. (color online) PXRD patterns and single crystal structure analysis of $LaO_{0.5}F_{0.5}Bi_{1-x}Pb_xS_2$. (a) PXRD pattern from $x$=0 to 0.10 on the $LaO_{0.5}F_{0.5}Bi_{1-x}Pb_xS_2$ and $NdO_{0.7}F_{0.3}Bi_{1-x}Pb_xS_2$. (b) Enlarged figure of Fig. 1(a) near (004) and (110) peaks. (c,d) Lattice parameters along the $a$, $c$-axis for $LaO_{0.5}F_{0.5}Bi_{1-x}Pb_xS_2$ and $NdO_{0.7}F_{0.3}Bi_{1-x}Pb_xS_2$.

Figure 1(a) shows powder X-ray diffraction patterns for $LaO_{0.5}F_{0.5}Bi_{1-x}Pb_xS_2$ ($x$=0-0.10). Since all the peaks correspond to the peaks of $CeOBiS_2$ type structure with the space group $P4/nmm$ symmetry, Pb substituted samples were successfully synthesized up to $x$=0.10. Figure 1(b) shows the magnified figure near (004) and (110) peaks of Fig. 1(a). (110) peaks is hardly changed by Pb substitution until $x$=0.1. On the other hand, (004) peak show a complex change by Pb substitution. To see this behavior in detail, lattice constants along $a$- and $c$-axis are plotted in Fig 1(c, d). Figure 1(c) shows the lattice constant along the $a$-axis as a function of Pb concentration. The lattice constant is almost constant up to $x$=0.06, whereas a slight increase is observed around $x$=0.09. The lattice constant along the $c$-axis is also constant up to $x$=0.06. In contrast to the $a$-axis, a steep decrease in the lattice constant is observed around $x$=0.09 as shown in Fig. 1 (d).

Pb substitution effect has been reported in $NdO_{0.7}F_{0.3}BiS_2$ [17]. In this compound, the Pb concentration dependence of the lattice constants was monotonic: the lattice constant along the c-axis decreases with increasing Pb concentration and the lattice constant along the $a$-axis increases, as shown in Fig 1(b,c). This behavior is contrasted with the case in $LaO_{0.5}F_{0.5}Bi_{1-x}Pb_xS_2$ where the lattice constant shows sudden change around $x$=0.09. In addition, The magnitude of the lattice constant change along the $c$-axis due to Pb substitution is about three times larger than that of $NdO_{0.7}F_{0.3}BiS_2$. These differences indicate that the Pb substitution effect in $LaO_{0.5}F_{0.5}BiS_2$ is qualitatively different from $NdO_{0.7}F_{0.3}BiS_2$. It is noted ionic $Pb^{2+}$ ion is generally larger than that of $Bi^{3+}$ ion. Therefore, the lattice constants are speculated to be enlarged by Pb substitution. In contrast to the specuration, the lattice constant along *c*-axis decreases with substituting Pb ion. So far, we don't know the reason why this decrease occurs. However, a decrease of the lattice constant along *c*-axis is observed in Pb substituted $NdO_{0.7}F_{0.3}BiS_2$ shown as blue square in Fig. 1(d). This tendency is considered to be common feature of Pb substitution.

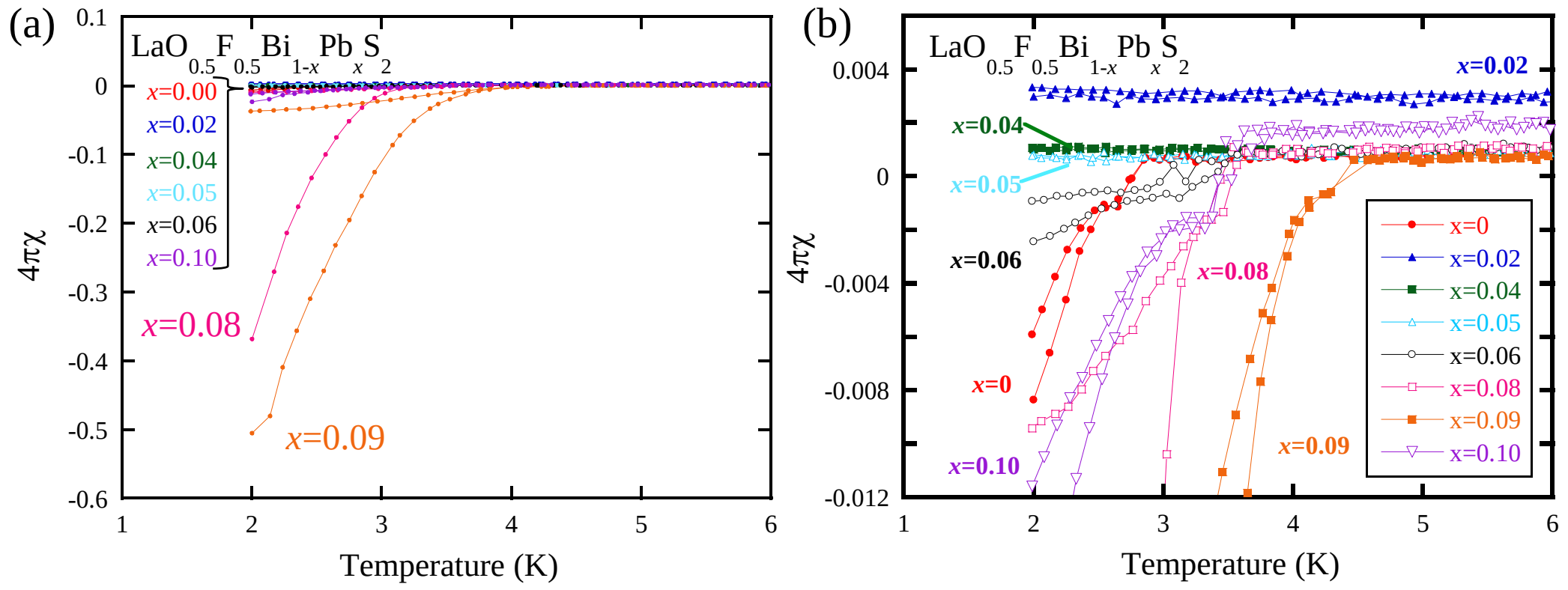

Figure 2. (Color online) Temperature dependence of magnetic susceptibility for $LaO_{0.5}F_{0.5}Bi_{1-x}Pb_xS_2$. (a) Temperature dependence of magnetic susceptibility for $LaO_{0.5}F_{0.5}Bi_{1-x}Pb_xS_2$ from 1K to 6K. (b) The magnified figure near the superconducting transition for $LaO_{0.5}F_{0.5}Bi_{1-x}Pb_xS_2$ .

Figure 2 (a) and (b) show the temperature dependence of magnetic susceptibility for $LaO_{0.5}F_{0.5}Bi_{1-x}Pb_xS_2$ ($x$=0-0.10) at magnetic field of 10 Oe. The sample at $x$=0 shows diamagnetic signal due to the appearance of superconductivity at around 3 K. The observed $T_c$ is consistent with previously reported one in polycrystalline samples, whereas the superconducting volume fraction in the single crystal obtained is smaller than that of the polycrystals [2]. Between $x$=0.02 and $x$=0.05, magnetic susceptibility does not show diamagnetic signal, indicating $T_c$ is reduced below 2 K. However, further increase in $x$ restores superconductivity above 2 K: Above $x$=0.06, $T_c$ becomes about 5 K, which is higher than that at $x$=0. Simultaneously, the superconducting volume fraction at $x$=0.08 and 0.09 are largely enhanced up to approximately 50 %. This enhancement of the superconductivity is achieved in only small range between $x$=0.08 and 0.09. At $x$ = 0.10, $T_c$ and superconducting volume fraction are reduced again. Thus, the enhancement of $T_c$ and superconducting volume fraction by Pb substitution in $LaO_{0.5}F_{0.5}Bi_{1-x}Pb_xS_2$ ($x$=0-0.10) is not monotonic.

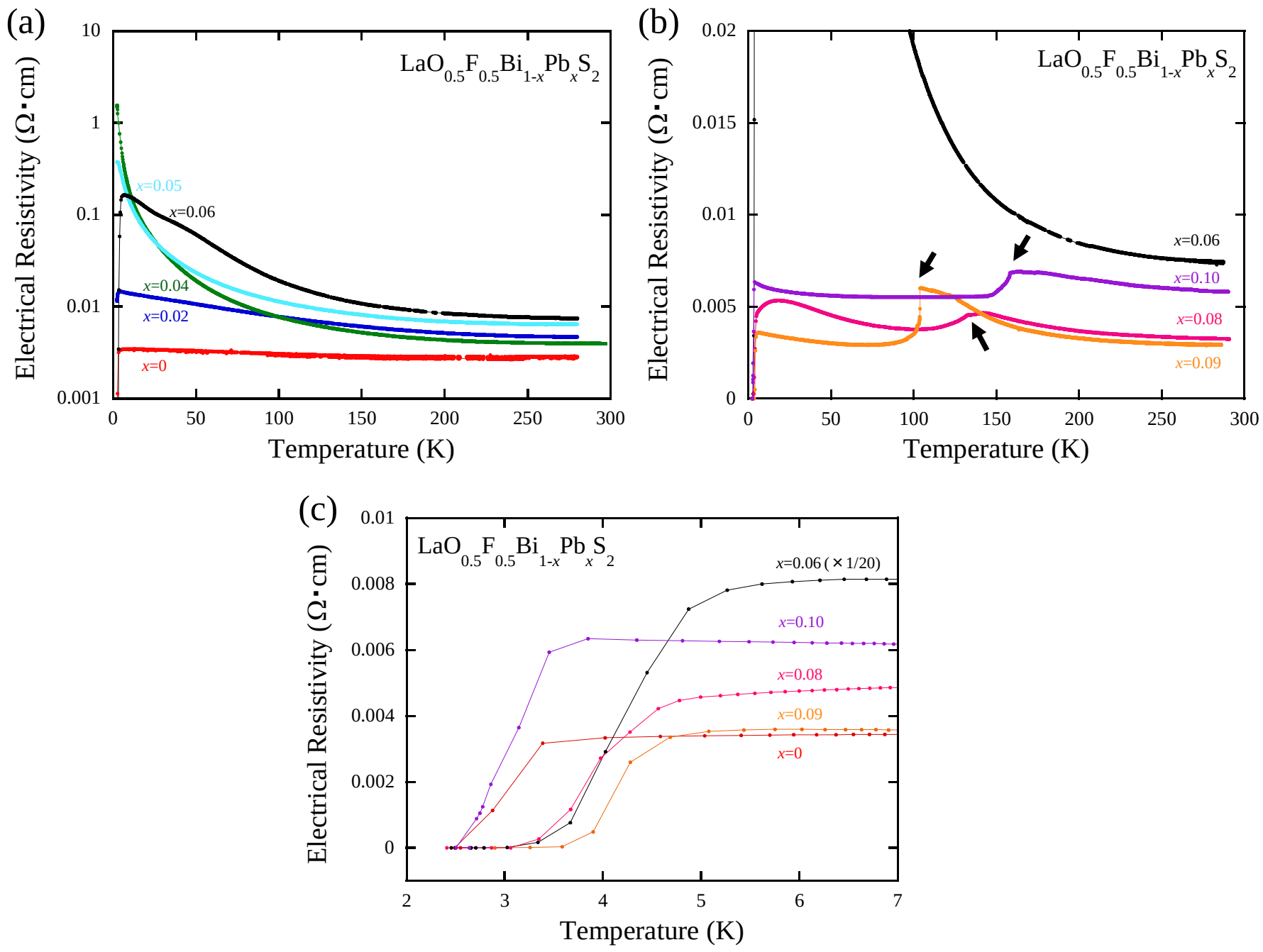

Figure 3. (Color online) Temperature dependence of electrical resistivity for $LaO_{0.5}F_{0.5}Bi_{1-x}Pb_xS_2$.
(a) Temperature dependence of electrical resistivity for $LaO_{0.5}F_{0.5}Bi_{1-x}Pb_xS_2$ (x=0.00-0.06) from 2K to 300K plotted in a logarithmic scale. (b) Temperature dependence of electrical resistivity for $LaO_{0.5}F_{0.5}Bi_{1-x}Pb_xS_2$ (x=0.06-0.10) from 2K to 300K. (c) The electrical resistivity near the superconducting transition.

Figure 3 (c) shows the temperature dependence of electric resistivity near $T_c$. Superconducting transition around 3 K is observed at the sample at $x$=0. Between $x$=0.02 and $x$=0.05, samples do not show the superconducting transition down to 2 K (not shown). However, superconducting transition appears again above $x$=0.05. Onset of the transition becomes about 5 K.

Figure 4 (a) summarizes the observed $T_c$. Here, $T_c^{Magnet}$ is defined as a temperature at which magnetic susceptibility begins to decrease, $T_c^{onset}$ is defined as a temperature at which resistivity begins to decrease, and $T_c^{zero}$ is defined as the temperature where resistivity becomes zero. The transition temperature determined by resistivity measurements well correlate to those determined by the magnetic measurement. It is apparent that observed Pb concentration dependence of the transition temperature is neither monotonic nor dome like behavior obtained by carrier doping in an electronic phase diagram where superconductivity appears rather wide carrier range.

Figure 4 (b) shows the superconducting volume fraction ($-4\pi\chi$). Again, Pb concentration dependence of the volume fraction is neither monotonic nor dome like: Although the volume fraction at $x$=0 is small, the increase in the superconducting volume fraction is in narrow range of $x$ near $x$=0.09 conspicuous.

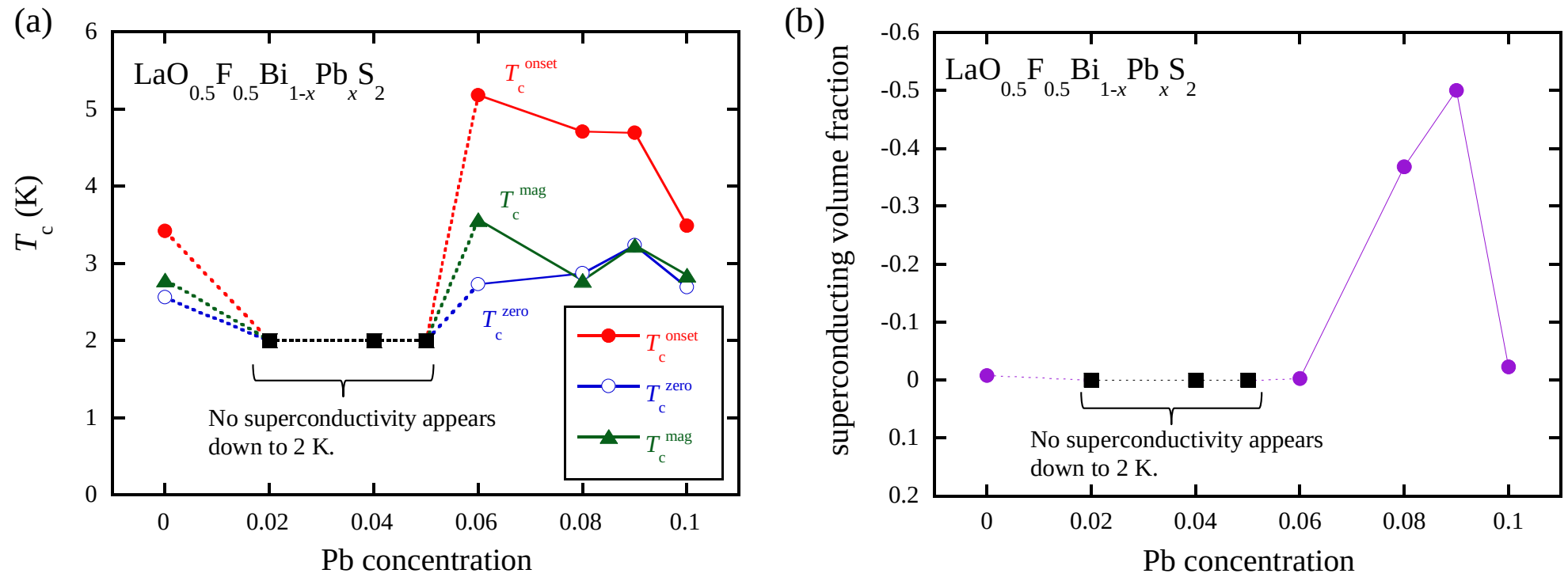

Figure 4. (a) Pb concentration $x$ dependence of $T_c$ for $LaO_{0.5}F_{0.5}Bi_{1-x}Pb_xS_2$. (b) Pb concentration $x$ dependence of superconducting volume fraction for $LaO_{0.5}F_{0.5}Bi_{1-x}Pb_xS_2$.

In $BiS_2$ materials, $T_c$ changes with carrier concentration. Because valence of Pb ion is typically +2 and that of Bi ion is typically +3, the electron carrier number is reduced with Pb substitution: 0.1 hole per Bi ion in maximum is doped in the sample at $x$=0.1. This carrier change may cause the change in $T_c$. However, this is not the case. Reported carrier dependence of $T_c$ in single crystals of $LaO_{1-y}F_yBiS_2$ shows that carrier doping about 0.1 hole per Bi ion does not change $T_c$ significantly around $y$=0.5[19,20]. In addition, hole doping tends to reduce $T_c$, which is inconsistent with the observed enhancement of $T_c$ around $x$=0.09. Thus, the observed enhancement of the superconducting properties cannot be explained by the change in the carrier number. As can be seen from Fig. 4 (a), $T_c$ drops at $x$=1.0. $T_c$ at $x$=1.0 is close to that at $x$=0. On the other hand, as shown in Fig.1 (c), the lattice constant along the $c$-axis drops steeply at $x$=0.8 and 0.9, then recovers the value close to that at $x$=0. Thus, the drop of $T_c$ at $x$=1.0 clearly corresponds to the change in the lattice constant along the $c$-axis. This indicates the change in $T_c$ is related to the change in an atomic configuration.

In addition to the enhancement of superconducting properties around $x$=0.09, the temperature dependence of resistivity shows anomalous behavior in this $x$ range. Figure 3 (a) and (b) show the temperature dependence of the electric resistivity for $LaO_{0.5}F_{0.5}Bi_{1-x}Pb_xS_2$ ($x$=0-0.10) from 2 K to 300 K. At $x$=0, resistivity shows very weak temperature dependence. With Pb substitution, the temperature dependence exhibits semiconducting behavior and shows steep increase at low temperature at $x$=0.04 and $x$=0.05 as shown in Fig. 3 (a). However, above $x$=0.06, where superconductivity is enhanced, the semiconducting behavior becomes weaker. Rather flat temperature dependence realizes above $x$=0.08 as shown in Fig. 3 (b). Furthermore, anomalous jump in resistivity appears around 150K, shown as arrows in Fig 3(b). Especially at $x$=0.09, at which $T_c$ becomes maximum, the resistivity jump is clear and discontinuous. Such anomalies in the resistivity are observed only above $x$=0.08, though shoulder like behavior can be seen at 50 K at $x$=0.06. Anomalous behavior in the temperature dependence of resistivity has been observed in $BiS_2$-based and $BiSe_2$-based superconductors [25-27]. However, clear and discontinuous jump observed in this study has not been reported. In addition, the appearance of this anomaly can be managed by Pb substitution. Thus, the observed anomaly in the resistivity is due to Pb substitution effect.

We observed Pb substitution effects on $LaO_{0.5}F_{0.5}BiS_2$. The reason why such Pb substitution effect was observed in $LaO_{0.5}F_{0.5}BiS_2$, not in $NdO_{0.7}F_{0.3}BiS_2$ is thought to be the difference in the structural stability: $LaO_{0.5}F_{0.5}BiS_2$ shows high $T_c$ phase at about 0.5 GPa, whereas $NdO_{0.5}F_{0.5}BiS_2$ changes to high $T_c$ phase at about 1.8 GPa [30, 31]. This seems to indicate that La system is more unstable than Nd system against pressure, although the O to F ratio is different from that of the sample used in this report. On the other hand, with Pb substitution in $LaO_{0.5}F_{0.5}BiS_2$, the superconducting properties are enhanced in rather narrow Pb concentration range after the suppression of $T_c$. The lattice constant along the $c$-axis shows steep drops at the Pb concentration where the superconducting properties are enhanced. Furthermore, at this concentration, the electrical resistivity shows discontinuous jump around 150 K. These results indicate that the Pb substitution effect in $LaO_{0.5}F_{0.5}BiS_2$ is qualitatively different from chemical pressure effect reported in some $BiS_2$ compounds where the change of $T_c$ is rather gradual. As we mentioned earlier, La system is more unstable than Nd system against pressure. Furthermore, the change in $T_c$ clearly corresponds to the change in the lattice constant along the *c*-axis. Thus, the observed anomalies seem to stem from the structural instability against external perturbations in $LaO_{0.5}F_{0.5}BiS_2$. In $BiS_2$ system, many phases or appearance of CDW have been reported [15]. Recently, supermodulation with a same signature of the predicted CDW is observed in $LaO_{0.5}F_{0.5}BiSe_2$ although this conduction layer is different from $BiS_2$ plane [32]. Unfortunately, because of the lack of the structural analysis at low temperature, the structure of $LaO_{0.5}F_{0.5}Bi_{1-x}Pb_xS_2$ has not determined yet. To confirm the low temperature state in $LaO_{0.5}F_{0.5}Bi_{1-x}Pb_xS_2$ ($x$=0.08, 0.09), detailed structural analysis such as crystal structural analysis or scanning tunnelling microscopy measurements are expected.

## 3. Summary

We examined Pb substitution effect in a single crystal of a layered superconductor $LaO_{0.5}F_{0.5}BiS_2$. The substitution of Pb for Bi at around $x$=0.08 and 0.09 enhanced superconducting temperature and the superconducting volume fraction. In this narrow region of Pb concentration, the lattice constants showed a deviation from continuous change and anomalous resistivity jump. These results seem to stem from the structural instability against external perturbations in $LaO_{0.5}F_{0.5}BiS_2$. These results give us new way to improve superconducting properties in $BiS_2$-based superconductors.

## Acknowledgement

This work was partly supported by a Grant-in-Aid for Young Scientists (B) (No. 15K17710).